\documentclass[12pt]{article}
\usepackage{graphicx}
\usepackage{epsfig}
\usepackage{amsbsy}

\begin{document}
\baselineskip 10mm

\centerline{\large \bf Thermal stability of cubane C$_8$H$_8$}

\vskip 6mm

\centerline{M. M. Maslov, D. A. Lobanov, A. I. Podlivaev, and L. A. Openov$^{*}$}

\vskip 4mm

\centerline{\it Moscow Engineering Physics Institute (State
University), 115409 Moscow, Russia}

\vskip 2mm

$^{*}$ E-mail: LAOpenov@mephi.ru

\vskip 8mm

\centerline{\bf ABSTRACT}

The reasons for the anomalously high thermal stability of cubane C$_8$H$_8$
and the mechanisms of its decomposition are studied by numerically simulating the dynamics of this metastable cluster at
$T=$ 1050 - 2000 K using a tight-binding potential. The decomposition activation energy is found from the temperature
dependence of the cubane lifetime obtained from the numerical experiment; this energy is fairly high,
$E_a=1.9\pm 0.1$ eV. The decomposition products are, as a rule, either C$_6$H$_6$ and C$_2$H$_2$ molecules or the isomer
C$_8$H$_8$ with a lower energy.

\newpage

Cubane C$_8$H$_8$ (Fig. 1) discovered in 1964 [1] is of great interest from both the fundamental and practical
standpoints. In this cluster, the carbon atoms are in cube vertices; i.e., the angles between the C-C covalent
bonds are 90$^0$ rather than 109.5$^0$ as in carbon compounds with a tetrahedral atomic arrangement and $sp^3$
hybridization of the atomic orbitals. Such large bending of the C-C-C bonds is energetically unfavorable. However,
hydrogen atoms arranged on the main diagonals of the cube stabilize this atomic configuration (Fig. 1),
corresponding to a local rather than global minimum of the potential energy as a function of the atomic coordinates.
Although cubane is a metastable cluster, its high stability is demonstrated by the experimental fact that
the cubane molecules not only retain their structure at temperatures significantly higher than room temperature
but also can form a molecular crystal, namely, solid cubane s-C$_8$H$_8$, with a melting temperature near 400 K
[2]. The formation heat of cubane is relatively large, 6.5 eV/C$_8$H$_8$ [3] (such energy is released, e.g., when
solid cubane transforms into graphite layers and H$_2$ molecules). High energy content of cubane makes it a
promising material for fuel elements, and the possibility of replacing the hydrogen atoms with various functional
groups (such as CH$_3$ in methylcubane [4]) opens the way to the synthesis of new compounds with unique
properties.

A wide application of cubane is hampered by the absence of cheap methods for its mass production [5].
In our opinion, detailed studies of the mechanisms and products of cubane decomposition may suggest a direction
of searching for new methods of its production.

In this case, it is interesting to consider a possible
reversal of the chemical reaction (e.g., under heating in the presence of appropriate catalysts lowering the barrier
of the reverse reaction). A certain analogy may be drawn with fullerene C$_{60}$: its decomposition is preceded
by a series of Stone-Wales transformations, which result in the formation of "surface" defects and, eventually,
a C$_2$ dimer is separated; on the other hand, annealing of these defects leads (through the same transformations)
to the formation of a fullerene from a strongly distorted spherical cluster C$_{60}$ (see [6, 7] and references
therein). There are only a few experimental studies of cubane decomposition (see, e.g., [8, 9]), and those
experiments were performed only over narrow temperature [8] and lifetime [9] ranges. As for theory, many
theoretical studies employed the same schematic potential-energy surface of cubane and its isomers [8], while,
to the best of our knowledge, the cubane dynamics before the instant of transition to another isomer has
been studied only at very high temperatures and for a very short time ($\sim 1$  ps) corresponding to only several
tens of cluster oscillation periods [10].

The main aim of this work is to numerically simulate the cubane dynamics over a wide temperature
range and determine the cubane decomposition activation energy, products of its decomposition, and the
types of isomers forming at the stage of evolution preceding the decomposition. We calculated the energies
of various atomic configurations within a nonorthogonal tight-binding model that was proposed for hydrocarbon
compounds in [11] and modified in [12] using a criterion of more exact correspondence between the
theoretical and experimental values of the binding energies and interatomic distances in various C$_n$H$_m$
molecules. The model is a reasonable compromise between more rigorous {\it ab initio} approaches and extremely simplified
classical potentials of interatomic interaction. In this model, the bond lengths in cubane are calculated to
be $l_{C-C}=1.570$ \AA~and $l_{C-H}=1.082$ \AA, which are close to experimental values 1.571 and 1.097 \AA, respectively
[13]. The calculated binding energy of the atoms in cubane $E_b=[8E(C)+8E(H)-E(C_8H_8)]/16=4.42$ eV/atom likewise
agrees with the experimental value 4.47 eV/atom [13]. The ratio of the energies of the carbon and hydrogen subsystems
in the heat-insulated cubane calculated in this model coincides with the
theoretical value [14].

We studied the thermal stability of cubane C$_8$H$_8$ by the molecular-dynamics method. At the initial instant
of time, random velocities and displacements are given to each of the atoms in such a manner that the momentum
and the angular momentum of the whole cluster are equal to zero. Then, the forces are calculated acting on
the atoms. The classical Newton equations of motion are numerically integrated using the velocity Verlet
method. The time step was $t_0=2.72\cdot 10^{-16}$ s. In the course of the simulation, the total energy of cubane (the
sum of the potential and kinetic energies) remained unchanged, which corresponds to a microcanonical
ensemble (the system is not in a thermal equilibrium with the surroundings [15-18]). In this case, the
"dynamic temperature" $T$ is a measure of the energy of relative motion of the atoms and is calculated from the
formula [19, 20] $\langle E_{kin} \rangle=\frac{1}{2}k_{B}T(3n-6)$, where $\langle E_{kin} \rangle$
is the time-averaged kinetic energy of the cluster, $k_B$ is the Boltzmann constant, and $n=16$ is the number of
atoms in cubane. It should be noted that the velocity Verlet algorithm is conservative with respect to the
momentum and the angular momentum [21] and the relative change in the total energy of cubane does not
exceed 10$^{-4}$ for at least $2\cdot 10^9$ molecular-dynamic steps, which corresponds to a time of $\sim 1 \mu$s.

We studied the cubane evolution for $\approx 50$ various sets of initial velocities and displacements of the atoms
corresponding to temperatures $T=1050-2000$ K. It turned out that, during its decomposition, cubane is most often
(in $\approx 80\%$ of events) transformed to the isomer cyclooctatetraene (COT) (Fig. 2a) with a lower potential energy
(a higher binding energy $E_b=4.82$ eV/atom). The cubane decomposition into a benzene C$_6$H$_6$ molecule
($E_b=4.82$ eV/atom) and an acetylene C$_2$H$_2$ molecule ($E_b=4.54$ eV/atom) occurs more rarely (in
$\approx 20\%$ of events) (Fig. 2b). We also observed several times the formation of styrene and some other isomers of
C$_8$H$_8$. Almost without exception, upon decomposition, cubane is transformed first into the isomer
syn-tricyclooctadiene (STCO) with $E_b=4.47$ eV/atom (Fig. 3a), which is quickly transformed (in a time of 0.1 - 1 ps)
into either COT or isomer bicyclooctatriene (BCT) with $E_b=4.65$ eV/atom (Fig. 3b). BCT, in turn, transforms
into COT or decomposes into benzene and acetylene (BEN + A) molecules. We have never observed reverse transitions,
such as STCO $\rightarrow$ cubane, COT $\rightarrow$ STCO, and BCT $\rightarrow$ STCO, whereas sometimes
COT is transformed into BCT with subsequent decomposition BCT $\rightarrow$ BEN + A.

As the temperature $T$ decreases from $\approx 2000$ to $\approx 1000$ K, the cubane lifetime
$\tau$ increases by six orders of magnitude, from $\sim 1$ ps to $\sim 1 \mu$s (Fig. 4). Since the decomposition of metastable
clusters is an inherently probabilistic process, the lifetime $\tau$ exhibits a dispersion at a given temperature
$T$. Nevertheless, it is seen from Fig. 4 that the results of the numerical simulation are described by the common
Arrhenius formula
\begin{equation}\label{1}
\tau^{-1}(T)=A\cdot\exp\left[-\frac{E_{a}}{k_{B}T}\right].
\end{equation}
According to this formula, the dependence of $ln(\tau)$ on $1/T$ is a straight line, whose slope determines the activation
energy $E_{a}=(1.9\pm 0.1)$ eV and its intersection point with the ordinate axis determines the frequency
factor $A=10^{16.03\pm0.36}$ s$^{-1}$. It is remarkable that the values of $E_a$ and $A$ agree well with the experimental
values $E_{a}=(1.87\pm0.04)$ eV and $A=10^{14.68\pm0.44}$ s$^{-1}$ obtained when studying cubane pyrolysis in a very narrow
range $T=(230\div260)$ $^{\circ}$C [8], which is far apart from the temperature range studied in this paper. The small
(on the logarithmic scale) difference in the frequency factors is likely due to the temperature dependence of $A$ (we note
that the value of $A$ for cubane is almost four orders of magnitude smaller than that for fullerene C$_{60}$ [7]).

Extrapolation of the $\tau(T)$ curve to the range $T < 1000$ K (which is inaccessible in direct numerical calculations
because of extremely long simulation time) permits one to compare the results of the simulation with the experimental values
of $\tau $ obtained in [9] for several temperatures in the range $T=373\div 973$ K. As seen from Fig. 4, here also there is
agreement between theory and experiment. Thus, Eq. (1) with the values of $E_a$ and $A$ found can be used to determine the
cubane lifetime (or, in any case, to make order-of-magnitude estimates) at both very high and comparatively low
temperatures. In particular, Eq. (1) gives $\tau\sim10^{16}$ s at room temperature and $\tau\sim10^{8}$ s at the melting
temperature of solid cubane $T_m\approx 400$ K (at which, on melting, only weak van der Waals bonds between C$_8$H$_8$
clusters are broken, whereas the clusters themselves retain their structure and the energy stored in them). The lifetime
decreases to $\tau\sim1$ s only on heating to $T\approx600$. Pyrolysis experiments [9] give $\tau\approx10$ ms at
$T=573$ K and $\tau\approx2$ ms at $T=673$ K (Fig. 4).

Note that, when analyzing the $\tau(T)$ dependence, we used Eq. (1) without a thermal-reservoir finite-size correction
[22, 23]. This correction reduces to replacing $T$ by $T-E_{a}/2C$ in the exponent of Eq. (1), where $C$ is the
microcanonical heat capacity of the cluster. With $C=k_{B}(3n-6)$, where $n=16$ is the number of atoms in
cubane, the closest fit between the modified Arrhenius formula and the numerical-simulation data is achieved
at $E_{a}=(1.41\pm 0.07)$ eV, which differs substantially from the experimental value [8] and, as we will see below, is
lower than the height $U$ of the minimum energy barrier to the cubane decomposition. The reason for this difference
is unclear, since earlier the inclusion of this correction allowed us to describe the experimental data on
the fullerene C$_{60}$ fragmentation [7]. Unlike fullerenes C$_{20}$ and C$_{60}$, cubane consists of unlike atoms and the
kinetic energy is nonuniformly distributed between the hydrogen and carbon subsystems in the thermally insulated
cubane [14]. This is likely to increase the effective heat capacity of cubane during its decomposition, and
the finite-size correction becomes insubstantial. However, this problem requires additional studies.

Let us find the height $U$ of the minimum energy barrier to cubane decomposition. Figure 5 presents calculated
energies of various isomers of cubane, decomposition products, and saddle points on the potential energy
hypersurface as functions of the atomic coordinates (the details of the calculation procedure can be
found in [15, 24, 25]). It is seen that the quantity $U$ is determined by the barrier to the transformation of
cubane into STCO isomer, which completely agrees with the molecular-dynamics simulation data. According
to our calculations, $U=1.59$ eV, which agrees with both the decomposition activation energy $E_{a}=(1.9\pm0.1)$ eV
(which we found from analyzing the numerical simulation data) and the experimental value $E_{a}=(1.87\pm0.04)$ eV [8].
As we might expect, the quantity $U$ is somewhat smaller than $E_a$, since in experiments (including numerical one) cubane
can decompose along paths with higher energy barriers. The fact that the barrier to the BCT $\rightarrow$ COT transition is
lower in height than that to decomposition of BCT into benzene and acetylene molecules explains why COT is a much
more frequent product of the cubane decomposition in numerical simulations.

We also calculated the quantity $U$ by the Hartree-Fock (HF) method without and with inclusion of the
Moller-Plesset second-order correction (MP2) and by the density functional method with the B3LYP exchange-correlation
functional. All the calculations are performed in the 6-31G* basis set. We found that $U=$ 3.95, 3.09, and 3.19 eV,
respectively. These values are much higher than the experimental value of $E_a$ [8] (despite the fact that, as noted above,
the inequality $U<E_a$ must take place). Thus, the results obtained in the tight-binding model agree much better with the
experimental data than the first-principle calculations do. This circumstance is due to the fact that we selected the
model parameters based on the requirement of the best agreement between the theoretical and experimental
characteristics of the various hydrocarbon molecules [12]. Note that one more decisive advantage of the
tight-binding model is the fact that the cluster evolution over a comparatively long time (on an atomic scale) of
$t\sim 1$ $\mu$s can be simulated (whereas for the {\it ab initio} methods $t\sim 1$ ps).

The results obtained in this work have allowed us to find the temperature dependence of the lifetime of the
metastable cubane C$_8$H$_8$, which may be useful in analyzing possible applications of cubane C$_8$H$_8$ and solid
cubane s-C$_8$H$_8$ as fuel elements. It is also interesting to develop new methods for synthesizing cubane based on
its known decomposition products.

\newpage
\includegraphics[width=\hsize,height=14cm]{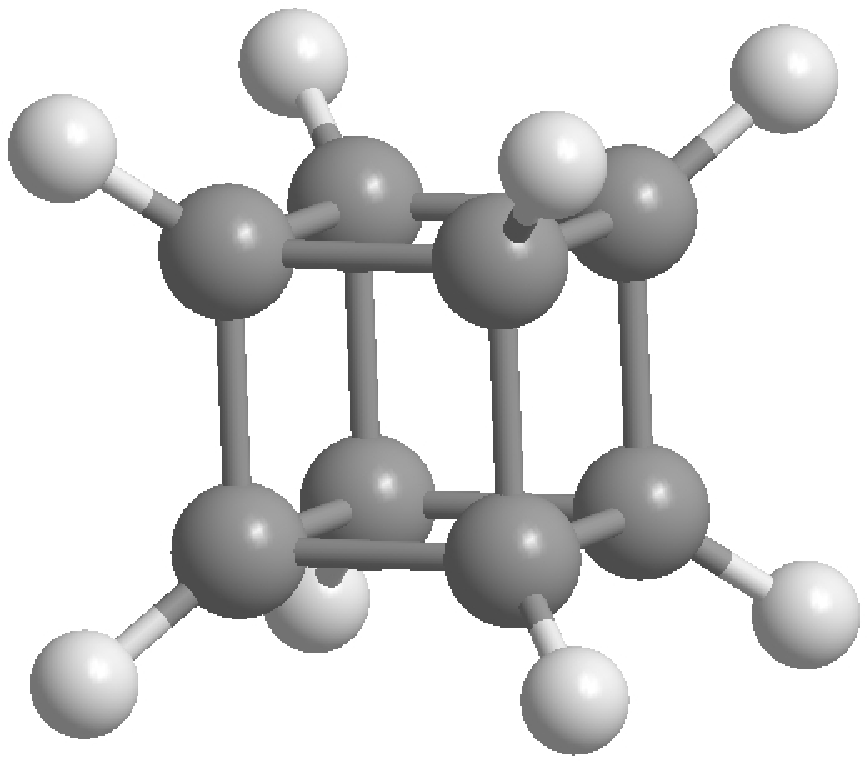}
\vskip 20mm
Fig. 1. Cubane C$_8$H$_8$ (schematic).

\newpage
\includegraphics[width=13cm,height=18cm]{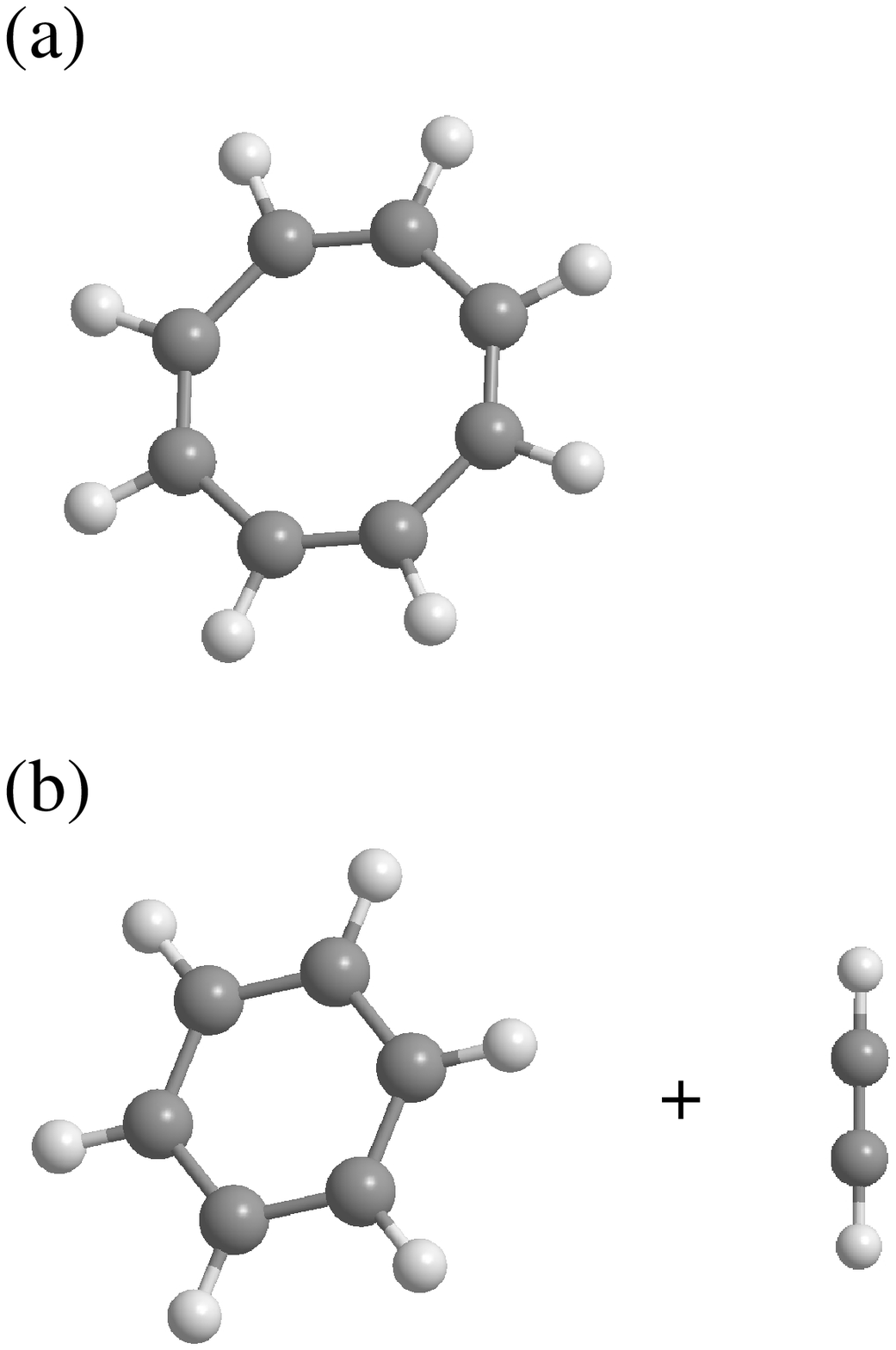}
\vskip 20mm
Fig. 2. Cubane decomposition products:
(a) isomer cyclooctatetraene (COT) and (b) C$_6$H$_6$ (benzene) and C$_2$H$_2$ (acetylene)
molecules.

\newpage
\includegraphics[width=10cm,height=18cm]{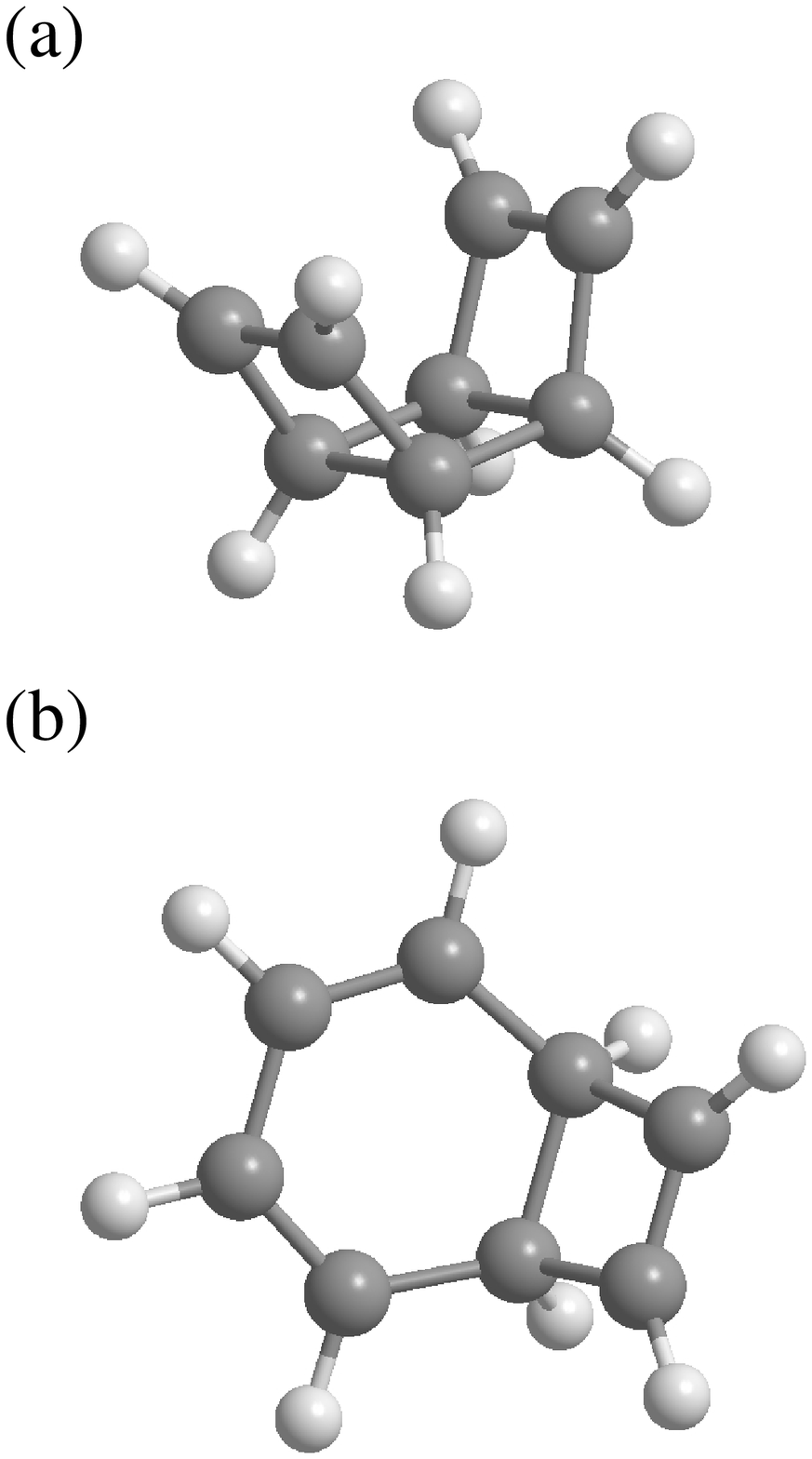}
\vskip 20mm
Fig. 3. Isomers of C$_8$H$_8$ that form at the cubane evolution stage preceding its decomposition: (a) isomer STCO (syn-tricyclooctadiene) and (b) BCT (bicyclooctatriene).

\newpage
\includegraphics[width=\hsize,height=15cm]{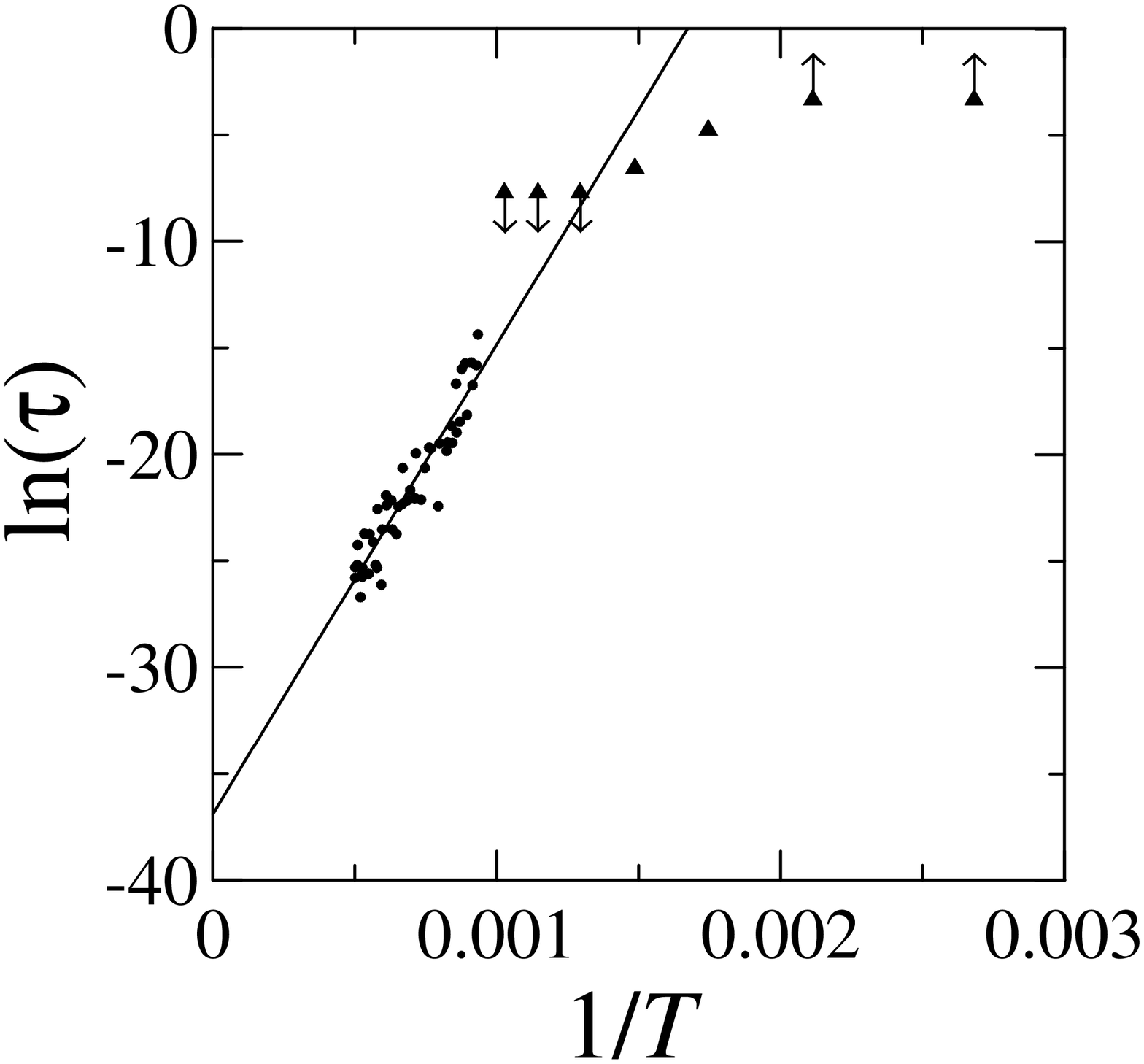}
\vskip 10mm
Fig. 4. Logarithm of the lifetime $\tau$ of cubane C$_8$H$_8$ plotted as a function of the reciprocal initial temperature $T^{-1}$:
circles are the results of calculations, the solid line is a least-squares linear fit, and triangles are the experimental data from [9]. The arrows indicate
the results obtained in [9] at temperatures $T\leq 474$ K and $T\geq 773$ K for which, because of technical problems, only
the lower ($\tau > 40$ ms) and upper ($\tau < 0.8$ ms) limitations on $\tau$, respectively, were determined.

\newpage
\includegraphics[width=\hsize,height=15cm]{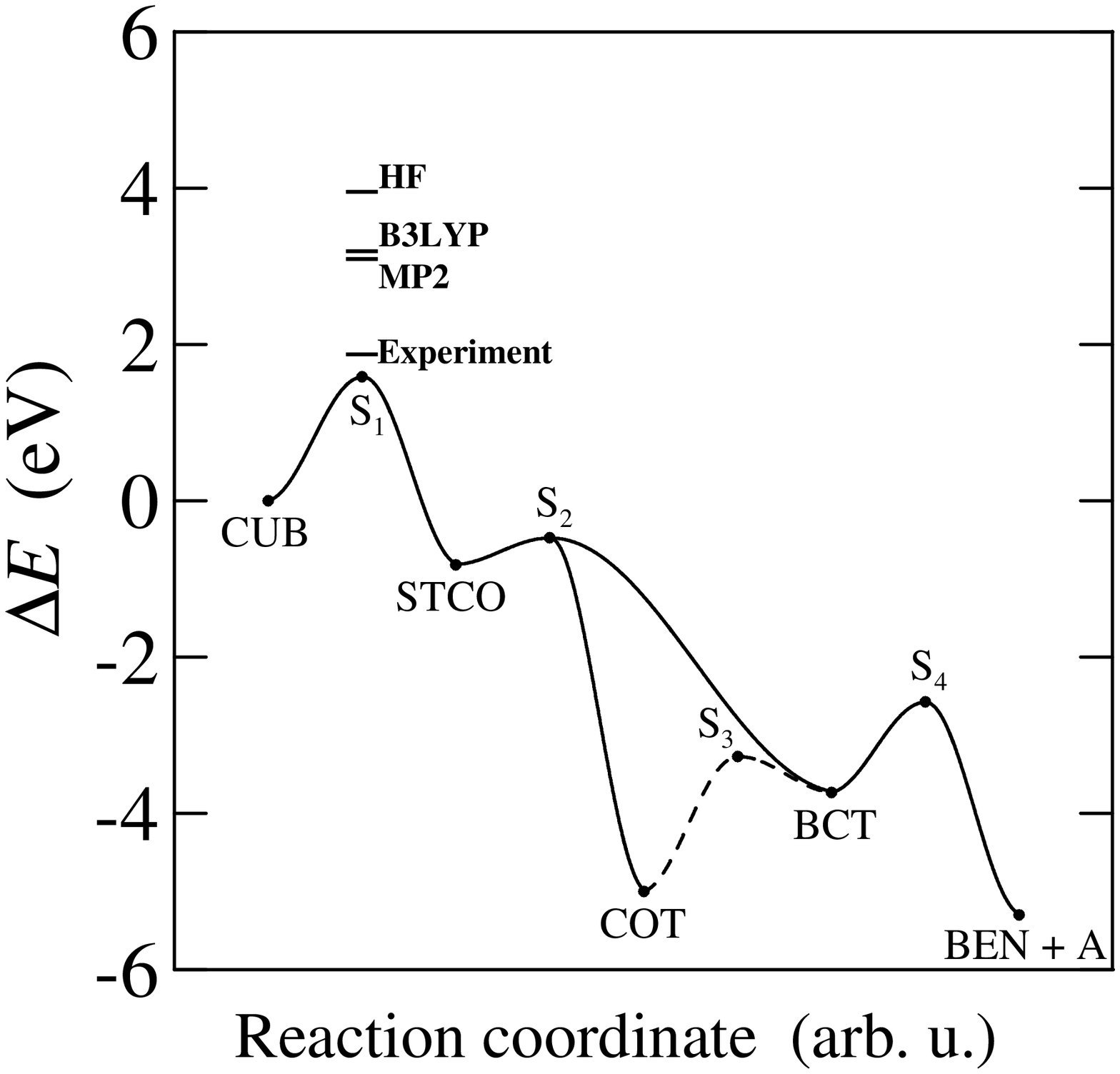}
\vskip 10mm
Fig. 5. Energies calculated by the tight-binding method for various isomers of cubane C$_8$H$_8$, its decomposition products,
and saddle points $S_i$ of the potential energy as a function of the atomic coordinates. The cubane energy is taken
as a reference point. The lines schematically show the paths of possible transitions. The experimental value of the cubane
decomposition activation energy [8] and the values of the minimum barrier to the transformation of cubane into STCO
calculated from first principles are also indicated (designations of isomers and {\it ab initio} methods are given in text).

\end{document}